\documentclass[11pt,preprint]{aastex}
\pdfoutput=1

\usepackage{graphics,graphicx}
\usepackage{wrapfig}

\begin{document}

\title{\LARGE{\bf Astrophysical Russian Dolls}}
\author{Abraham Loeb and Nia Imara}

\bigskip
\bigskip

{\it ~\\``As it unfolded, the structure of the story began to
  remind me of one of these Russian dolls\\ that contains innumerable
  ever-smaller dolls within.''}

\noindent
{\it Carlos Ruiz Zaf\'on, ``The Shadow of the Wind'' (2001)}

\bigskip

{\huge R}ussian dolls display a miniature doll embedded in the belly
of an identical version of itself, which lies in turn within an even
larger doll replica and so on. {\it Does the Universe exhibit Russian
  dolls?}  The immediate example which comes to mind is that electrons
move around nuclei within atoms that lie inside planets which orbit
around stars, as those stars circle around the center of the Milky Way
galaxy. Each of the ``dolls'' in this classic example attracts the
attention of a separate community of scientists which often ignores
the other ``dolls'' despite their similarities. Aside from the
aesthetic pleasure of recognizing scaled versions of similar systems,
drawing analogies between them may unravel a fundamental truth that
unifies their governing principles. The art of identifying common themes 
on different scales resembles the search for the common DNA characteristics of
relatives from the same family.

Are there other examples of astrophysical Russian dolls, and what
could we learn from their similarities? Below we list a few such
examples.

\begin{itemize}

\item {\bf Disks within disks.} Our Milky Way Galaxy consists of a
  disk of stars and gas, circling at a characteristic speed of 235
  kilometers per second around a common center.  At the Galactic
  Center lies a $4\times 10^6~M_\odot$ black hole around which swirls
  a circumnuclear disk of stars and gas.  Throughout the Galactic
  disk, newly forming stars, which are embedded in the molecular
  clouds circling the Galactic center, are also encircled by
  disks. The gas and dust in such protoplanetary disks eventually
  clump into planets, as was the case in our own Solar System five
  billion years ago.  But this may not be the final ``doll'' in this
  system of astrophysical disks.  State-of-the-art simulations suggest
  that planets, in the early stages of their evolution, are surrounded
  by a miniature disk of gas (see the smallest doll in Figure
  \ref{fig1}).  Future advancements in technology might enable us to
  detect the presence of such disks around nascent planets.

\begin{figure}[ht]
\epsscale{0.8}
\plotone{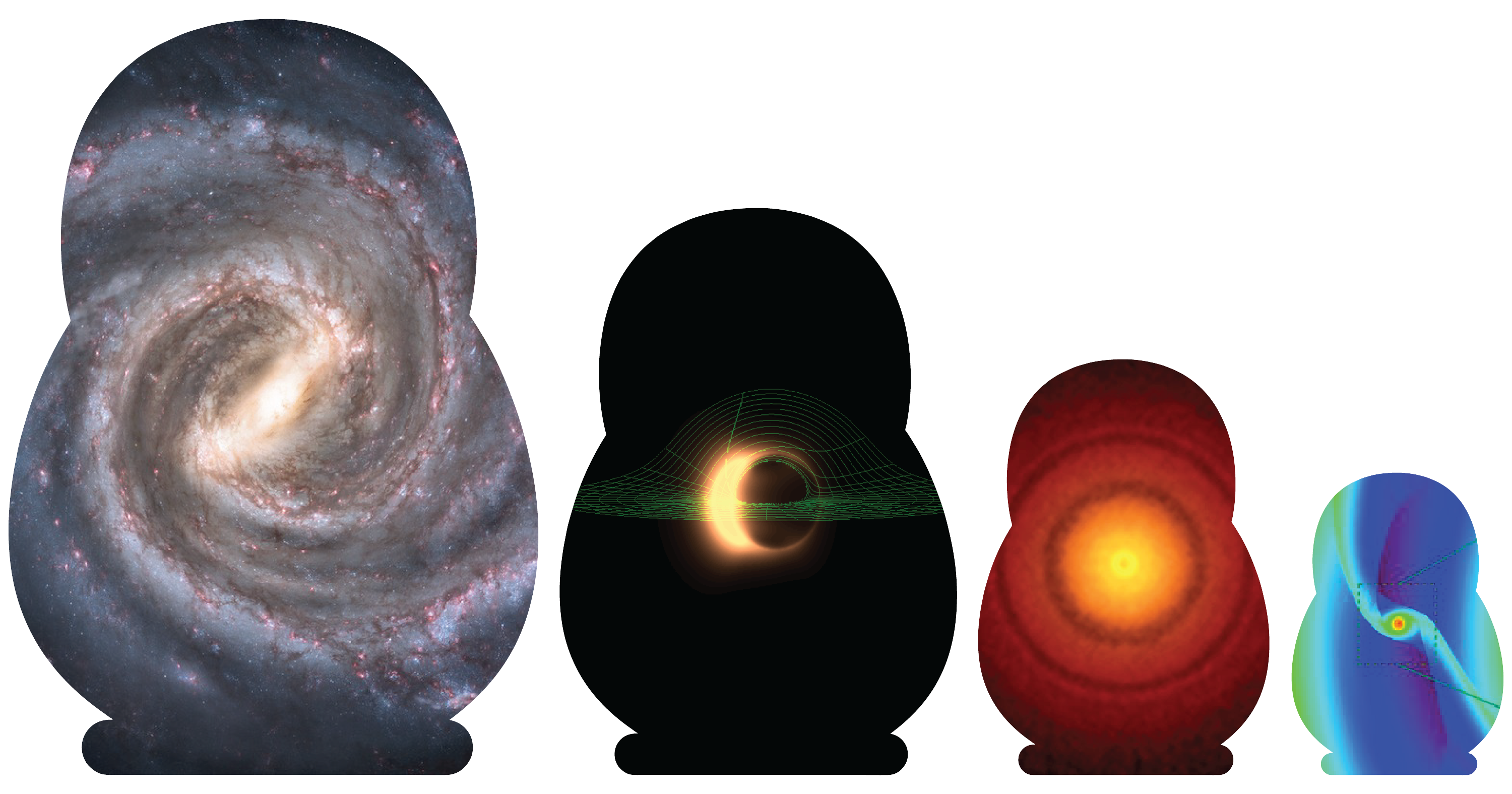}
\caption{Astrophysical Russian dolls---disks. Credits for dolls, largest to smallest: 
  Artist's impression of a spiral galaxy (NASA). Simulation of Sgr A$^\ast$ and its disk 
  (A.E. Broderick \& A. Loeb).  Observation of protoplanetary disk 
  (S. Andrews/B. Saxton/ALMA).  Simulation of a disk around a planet 
  (James Stone and collaborators, Princeton.)}
  \label{fig1}
\end{figure}

Asteroid formation in a disk around a planet and planet formation in a
disk around a star may have fundamental similarities to molecular cloud
formation in the Galactic disk and star formation around the central
black hole. By recognizing generic dynamical processes in one of these
systems, one could make new predictions for the properties of the
others.

Of course, in drawing such analogies one should keep in mind the
important differences between galactic and protoplanetary disks,
including the different temperature scale, magnetic field strength,
turbulence, and ionization state of the gas.

\begin{figure}[ht]
\epsscale{0.8}
\plotone{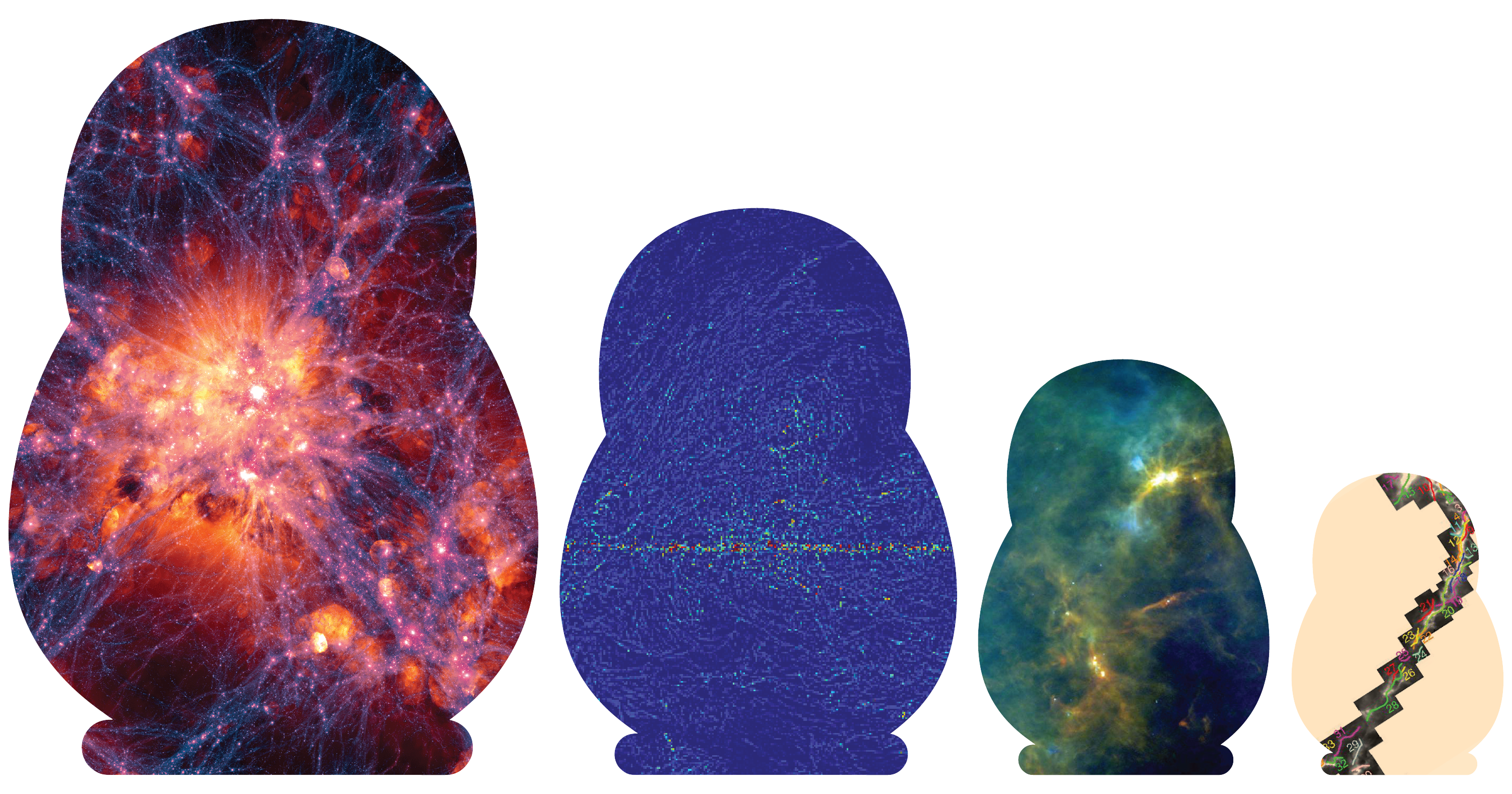}
\caption{Astrophysical Russian dolls---filaments.  Credits for dolls, largest to smallest:
  Simulation of intergalactic filaments (Lars Hernquist and collaborators, Harvard).
  Observation of Galactic neutral atomic hydrogen (P. M. W. Kalberla et al.).
  Observation of filaments in a molecular clouds (ESA/Herschel/PACS/SPIRE/V. Roccatagliata, U. M\"unchen).
  Observation of fibers within a filament (A. Hacar et al.).}
  \label{fig2}
\end{figure}

\item {\bf Filaments within filaments.} Under the action of its own gravity,
  each overdense region in the Universe tends to collapse first along
  its short axis - creating a sheet, then along its medium axis -
  creating a filament, and finally along its long axis - creating a
  compact object like a galaxy or a group of galaxies. As a result,
  the diffuse intergalactic medium (IGM) is organized into sheets and
  filaments, constituting a ``cosmic web'' (see Figure \ref{fig2})
  that serves as a skeleton for the large scale structures in the
  Universe.  Inside galaxies, which are located at the nodes of
  intersecting intergalactic filaments, the interstellar medium (ISM)
  repeats this pattern.  Within the Milky Way disk, for example, blast waves set
  off by supernovae may be responsible for the network of sheets and
  filaments characterizing the morphology of the interstellar atomic
  gas, as revealed by recent high-resolution observations.  Colliding
  flows of neutral atomic gas or other types of instabilities trigger
  the formation of molecular clouds, the dark, frigid structures
  inside of which stars form.  In recent years, infrared observations
  by the \emph{Herschel Space Telescope} have revealed that molecular
  clouds are threaded by complex webs of parsecs-long, skinny, dense
  filamentary structures.  Once the mass per unit length of a filament
  exceeds a critical value, it may become gravitationally unstable and
  fragment into pre-stellar cores.  Observations and simulations of
  interstellar filaments indicate that embedded cores grow by
  accreting gas channeled along these filaments.  Similarly, cosmological
  simulations suggest that cold streams of gas flowing along IGM filaments
  supply galaxies with the bulk of the fuel required for star
  formation.  Scaled threadlike versions of similar substructure may exist in this 
  system of filamentary Russian dolls.  Possible evidence for so-called 
  ``fibers'' have been observed in Galactic molecular
  clouds and in ISM simulations.  The leading interpretation is that
  large scale filaments in the ISM are not simple cylindrical
  structures but, rather, are composed of intricate bundles of fibers
  which, if they are gravitationally unstable, ultimately fragment
  into cores.

Despite the similarities in appearance, the formation mechanisms of
intergalactic and interstellar filaments may be quite different.  The
consensus view of cosmologists is that the IGM filaments grew out of
gravitational instability, whereas filaments inside molecular clouds
may arise from magneto-hydrodynamic turbulent compression of
interstellar gas. However, some fine structure of the IGM filaments
may be induced by outflows from galaxies which shape the gas around
them similarly to the way stellar feedback and turbulence shape the 
small-scale structure in the ISM.

\item{\bf Clusters within clusters.} Galaxies tend to cluster. And each
  spiral galaxy includes a gravitationally bound disk of gas,
  containing molecular clouds that are frequently clustered in
  giant molecular associations, which are possibly held
  together by the mutual gravitational attraction.  While the smallest
  molecular clouds in our Galaxy may be confined by the pressure of
  the ambient interstellar medium, it is widely accepted that the most
  massive, \emph{giant} molecular clouds are held together by gravity.
  Nested hierarchically within molecular clouds are dense cores of
  gas, the very densest of which go on to form clusters of stars.

Thus, the long range, scale free, force of gravity manifests itself in
similar ways over a wide range of clustering scales.

\end{itemize}

The pursuit of scientific knowledge is rooted not only in human
curiosity and the desire to understand the natural world, but also in
our innate need to enjoy and seek the beauty associated with the
patterns and symmetries of nature. Forging connections across
disciplinary borders enhances our perception of beauty, while
simultaneously leading to a more comprehensive understanding of the
Universe. The reinforcement acts also in reverse---as our
understanding of the Universe improves, so is our sense of its beauty.

\bigskip
\bigskip
\bigskip

\noindent{\it Abraham (Avi) Loeb and Nia Imara are at the
  Harvard-Smithsonian Center for Astrophysics, 60 Garden Street,
  Cambridge, Massachusetts 02138, USA. \\e-mail:
  aloeb@cfa.harvard.edu; nimara@cfa.harvard.edu}

\end{document}